\documentclass[aps,pre,twocolumn,showpacs,groupedaddress,amsmath,amssymb]%
{revtex4}

\usepackage{graphicx}
\usepackage{dcolumn}
\usepackage{bm}
\usepackage{subfigure}

\begin{document}

\title{Theory and simulation of two-dimensional nematic and tetratic phases}


\author{Jun Geng}
\author{Jonathan V. Selinger}
\email{jvs@lci.kent.edu}
\affiliation{Liquid Crystal Institute, Kent State University, Kent, OH 44242}


\date{\today}

\begin{abstract}
Recent experiments and simulations have shown that two-dimensional systems can
form tetratic phases with four-fold rotational symmetry, even if they are
composed of particles with only two-fold symmetry.  To understand this effect,
we propose a model for the statistical mechanics of particles with almost
four-fold symmetry, which is weakly broken down to two-fold.  We introduce a
coefficient $\kappa$ to characterize the symmetry breaking, and find that the
tetratic phase can still exist even up to a substantial value of $\kappa$.
Through a Landau expansion of the free energy, we calculate the mean-field
phase diagram, which is similar to the result of a previous hard-particle
excluded-volume model.  To verify our mean-field calculation, we develop a
Monte Carlo simulation of spins on a triangular lattice.  The results of the
simulation agree very well with the Landau theory.
\end{abstract}


\maketitle
\section{\label{sec:introduction}Introduction}
In statistical mechanics, one key issue is how the \emph{microscopic} symmetry
of particle shapes and interactions is related to the \emph{macroscopic}
symmetry of the phases.  This issue is especially important for liquid-crystal
science, where researchers control the orientational order of phases by
synthesizing molecules with rod-like, disk-like, bent-core, or other shapes.
In many cases, the low-temperature phase has the same symmetry as the
particles of which it is composed, while the high-temperature phase has a
higher symmetry.  For example, in two dimensions (2D), particles with a
rectangular or rod-like shape, which has two-fold rotational symmetry, form a
low-temperature nematic phase, which also has two-fold symmetry.  Likewise, if
the particles are perfect squares, which have four-fold rotational symmetry,
they can form a four-fold symmetric tetratic phase.

\begin{figure}[b]
\includegraphics{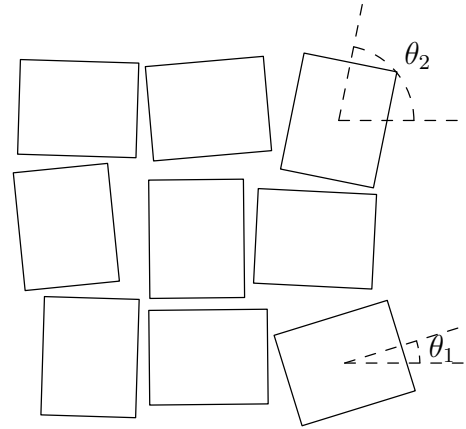}
\caption{\label{fig:model} Schematic illustration of an interacting particle
system in the tetratic phase. The shape of the particles indicates that the
rotational symmetry of the interaction is broken down from four-fold to
two-fold.}
\end{figure}

An interesting question is what happens if the symmetry of the particles is
slightly broken.  Will the symmetry of the phase also be broken, or can the
particles still form a higher-symmetry phase?  For example, we can consider
particles with approximate four-fold rotational symmetry that is slightly
broken down to two-fold, as in Fig.~\ref{fig:model}.  Can these particles
still form a tetratic phase, or will they only form a less symmetric nematic
phase?

Recently, several experimental and theoretical studies have addressed this
problem.  Narayan et al.~\cite{Narayan05} performed experiments on a
vibrated-rod monolayer, and found that two-fold symmetric rods can form a
four-fold symmetric tetratic phase over some range of packing fraction and
aspect ratio.  Zhao et al.~\cite{Nematicandal} studied experimentally the
phase behavior of colloidal rectangles, and found what they called an almost
tetratic phase.  Donev et al.~\cite{donev:054109} simulated the phase behavior
of a hard-rectangle system with an aspect ratio of 2, and showed they form a
tetratic phase.  Another simulation by Triplett et al.~\cite{triplett:011707}
showed similar results.  In further theoretical work, Mart\'{\i}nez-Rat\'{o}n
et al.~\cite{Martinez-Rat05,martinez-raton:011711} developed a
density-functional theory to study the effect of particle geometry on phase
transitions.  They found a range of the phase diagram in which the tetratic
phase can exist, as long as the shape is close enough to four-fold symmetric.
In all of these studies, the particles interact through hard, Onsager-like~
\cite{Onsager},
excluded-volume interactions.

The purpose of the current paper is to investigate whether the same phase
behavior occurs for particles with longer-range, soft interactions.  We
consider a general four-fold symmetric interaction, which is slightly broken
down to two-fold symmetry.  We first calculate the phase diagram using a
Maier-Saupe-like mean-field theory~\cite{MS1,MS2,MS3}.  To verify the theory,
we then perform
Monte Carlo simulations for the same interaction.

This work leads to two main results.  First, the tetratic phase still exists
up to a surprisingly high value of the microscopic symmetry breaking (as
characterized by the interaction parameter $\kappa$, which is defined below).
Second, the phase diagram is quite similar to that found by
Mart\'{\i}nez-Rat\'{o}n et al.\ for particles with excluded-volume repulsion.
This similarity indicates that the phase behavior is generic for particles
with almost-four-fold symmetry, independent of the specific interparticle
interaction.

The plan of this paper is as follows.  In Section~\ref{sec:model}, we present
our model and calculate the mean-field free energy.  We then examine the phase
behavior and calculate the phase diagram in Section~\ref{sec:r_model}.  In
Section \ref{sec:simulation} we describe the Monte Carlo simulation methods
and results.  Finally, in Section~\ref{sec:conclusions} we discuss and
summarize the conclusions of this study.

As an aside, we should mention one point of terminology.  The tetratic phase
has occasionally been called a ``biaxial'' phase, by analogy with 3D biaxial
nematic liquid crystals~\cite{Martinez-Rat05}.  However, this analogy is
somewhat misleading.  In 3D liquid crystals, the word ``biaxial'' refers to a
phase with orientational order in the long molecular axis and in the
transverse axes, i.e. a phase with \emph{lower} symmetry than a conventional
uniaxial nematic. By contrast, the tetratic phase has \emph{higher} symmetry
than a conventional nematic, four-fold rather than two-fold.  For that reason,
we will not use the term ``biaxial'' in this work.

\section{\label{sec:model}Model}

Maier-Saupe theory is a widely used form of mean-field theory, which describes
the isotropic-nematic transition in 3D liquid crystals.  In this section, we
extend Maier-Saupe theory to describe 2D liquid crystals with
almost-four-fold symmetry, as shown in Fig.~\ref{fig:model}.  For this
purpose, we use the modified Maier-Saupe interaction
\begin{equation}
U_{12}\left(r_{12},\theta_{12}\right)
=-U_0(r_{12})\left[\kappa\cos\left(2\theta_{12}\right)+
\cos\left(4\theta_{ 12}\right)\right]
\label{eq:interaction},
\end{equation}
where $\theta_{12}=\theta_1-\theta_2$ is the relative orientation angle
between particles 1 and 2, and $r_{12}$ is the distance between these
particles.  In this interaction, the dominant orientation-dependent term is
$\cos\left(4\theta_{12}\right)$, which has perfect four-fold symmetry.  The
term $\cos\left(2\theta_{12}\right)$ represents a correction to the
interaction, which has only two-fold symmetry.  If the coefficient $\kappa$ is
small, then the symmetry is slightly broken from four-fold down to two-fold.
(By contrast, if $\kappa$ is large, then the interaction clearly has two-fold
symmetry and the four-fold term is unimportant, as in classic Maier-Saupe
theory.)  The overall coefficient $U_0(r_{12})$ is an arbitrary
distance-dependent term.

In mean-field theory, we average the interaction energy to obtain an effective
single-particle potential due to all the other particles,
\begin{equation}
 U_{\textrm{eff}}\left(\theta_1\right)
=\int \mathrm{d}^2 {\bf r}_{12} \mathrm{d} \theta_2 \rho\left(\theta_2\right)
U_{12}\left(r_{12},\theta_{12}\right).
\label{eq:Ueff}
\end{equation}
Here, $\rho\left(\theta_2\right)$ is the orientational distribution function,
which is normalized as
\begin{equation}
\rho_0=\int_0^{\pi} \mathrm{d} \theta \rho\left(\theta\right),
\end{equation}
where $\rho_0$ is the number density of particles. To calculate
$U_{\textrm{eff}}$, we set the $x$-axis along an ordered direction (the
director in nematic case, or one of the two orthogonal ordered directions in
the tetratic case).  In that case, the averages of $\sin(2\theta)$ and
$\sin(4\theta)$ vanish by symmetry, and hence Eq.~(\ref{eq:Ueff}) becomes
\begin{equation}
U_{\textrm{eff}}\left(\theta\right)=-\bar{U}
\rho_0\left[\kappa C_2 \cos \left( 2\theta \right)
+C_4\cos\left(4\theta \right)\right],
\label{eq:Ueff2}
\end{equation}
where $\bar{U}$ is the integral over the position-dependent part of the
potential, and
\begin{subequations}
\label{eq:C2C4}
\begin{eqnarray}
C_2&=&\left<\cos\left(2\theta\right)\right>, \\
C_4&=&\left<\cos\left(4\theta\right)\right>.
\end{eqnarray}
\end{subequations}
The resulting orientational distribution function is
\begin{equation}
\rho\left(\theta\right)=
\frac{\rho_0\exp[\gamma(\kappa C_2\cos(2\theta)+C_4\cos(4\theta))]}%
{\int_0^{\pi}\mathrm{d}\theta
\exp[\gamma(\kappa C_2 \cos(2\theta)+C_4\cos(4\theta))]},
\end{equation}
where we have defined the dimensionless ratio $\gamma=\rho_0 \bar{U}/(k_BT)$.

Note that $C_2$ can be regarded as a nematic order parameter, and $C_4$ as a
tetratic order parameter.  In the isotropic phase, the system has $C_2=C_4=0$.
By comparison, in the tetratic phase, the system has $C_2=0$ but $C_4\not=0$.
In the nematic phase, with the most order, the system has $C_2\not=0$ and
$C_4\not=0$.

To determine which of these phases is most stable, we must calculate the free
energy $F=\left<U\right>+k_B T\left<\log\rho\right>$ as a function of the
order parameters $C_2$ and $C_4$.  The average interaction energy per particle
is
\begin{equation}
\left<U\right>=-\frac{1}{2}\bar{U}\rho_0\left(\kappa C_2^2+C_4^2\right).
\end{equation}
The entropic part of the free energy per particle is
\begin{eqnarray}
\lefteqn{k_B T\left<\log\rho\right>
=\bar{U}\rho_0\left(\kappa C_2^2+C_4^2\right)}\\
& & -k_B T \log\left(\frac{1}{\pi}\int_0^{\pi}\mathrm{d}\theta
\exp[\gamma(\kappa C_2 \cos2\theta +C_4\cos4\theta)]\right);\nonumber
\end{eqnarray}
here we have have subtracted off the constant entropy of the isotropic phase.
We combine these terms and normalize by $k_B T$ to obtain the dimensionless
free energy
\begin{eqnarray}
\label{eq:free}
\lefteqn{\frac{F}{k_B T}
=\frac{1}{2}\gamma\left(\kappa C_2^2+C_4^2\right)}\\
& & - \log\left(\frac{1}{\pi}\int_0^{\pi}\mathrm{d}\theta
\exp[\gamma(\kappa C_2 \cos2\theta +C_4\cos4\theta)]\right).\nonumber
\end{eqnarray}
Minimizing this free energy with respect to $C_2$ and $C_4$ gives the
equations
\begin{subequations}
\begin{eqnarray}
C_2&=&\frac{1}{\rho_0}\int_0^{\pi}\mathrm{d}\theta
\cos(2\theta)\rho\left(\theta\right),\\
C_4&=&\frac{1}{\rho_0}\int_0^{\pi}\mathrm{d}\theta
\cos(4\theta)\rho\left(\theta\right),
\end{eqnarray}
\end{subequations}
which are exactly consistent with Eqs.~(\ref{eq:C2C4}).

\begin{figure*}
(a)\subfigure{\includegraphics{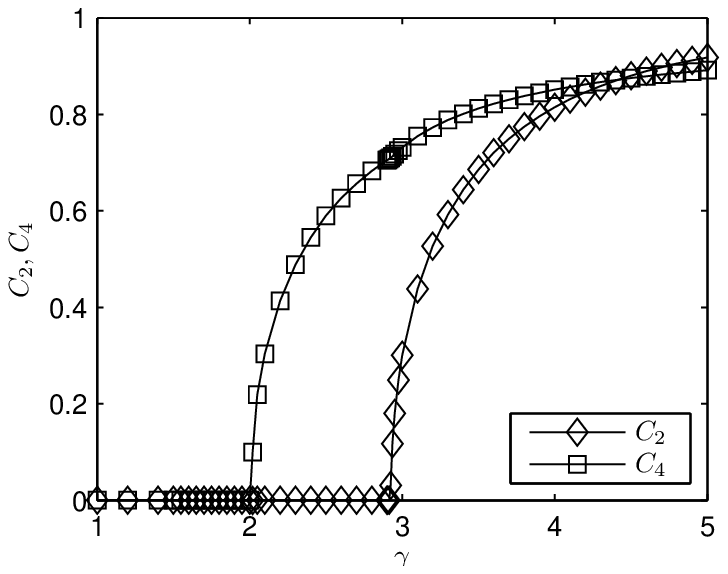}}
(b)\subfigure{\includegraphics{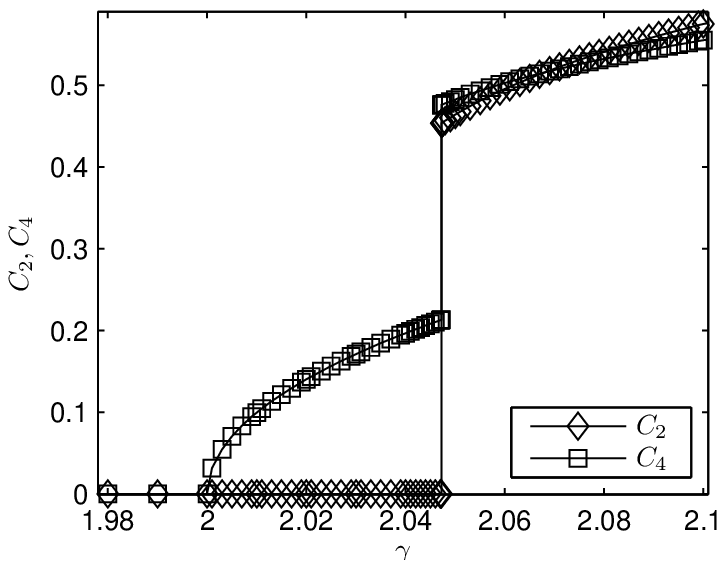}}
(c)\subfigure{\includegraphics{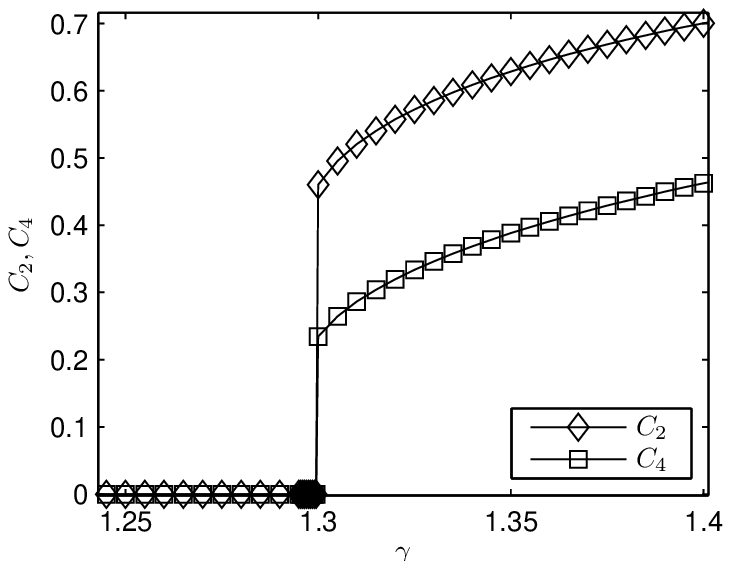}}
(d)\subfigure{\includegraphics{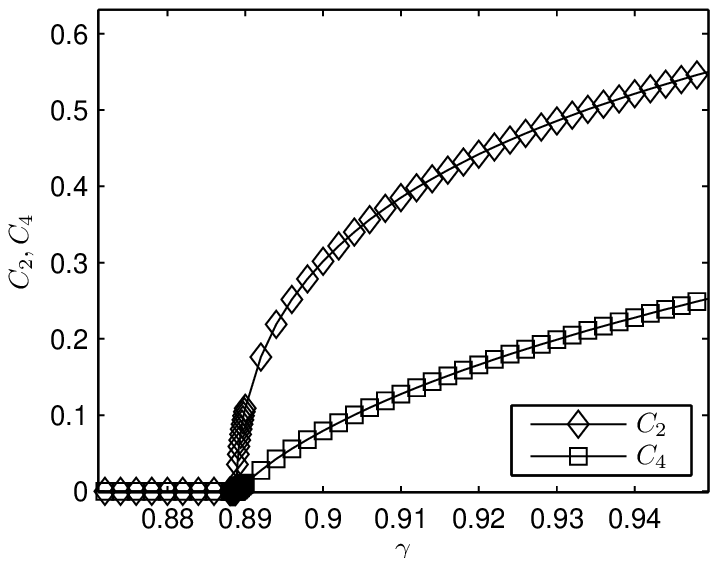}}
\caption{\label{fig:Csgkt} Numerical mean-field calculation of the order
parameters $C_2$ and $C_4$ as functions of $\gamma$ (inverse temperature), for
several values of $\kappa$ (two-fold distortion in the interaction):
(a)~$\kappa=0.4$.  (b)~$\kappa=0.75$.  (c)~$\kappa=1.5$.  (d)~$\kappa=2.25$.}
\end{figure*}

\section{\label{sec:r_model}Mean-field results}

The model is now completely defined by two dimensionless parameters:
$\gamma=\rho_0 \bar{U}/(k_BT)$ is the ratio of interaction energy to
temperature, and $\kappa$ represents the breaking of four-fold symmetry in the
interparticle interaction.  We would like to determine the phase diagram in
terms of these two parameters.  As a first step, we minimize the free energy
of Eq.~(\ref{eq:free}) numerically using Mathematica.  We then do analytic
calculations to obtain exact values for second-order transitions and special
points in the phase diagram.

Figure~\ref{fig:Csgkt} shows the numerical mean-field results for the order
parameters $C_2$ and $C_4$ as functions of $\gamma$, for several values of
$\kappa$.  These plots represent experiments in which the temperature is
varied, for particles with a fixed interaction.  When the four-fold symmetry
is only slightly broken by the small value $\kappa=0.4$, there are two
second-order transitions, first from the high-temperature isotropic phase to
the intermediate tetratic phase, and then from the tetratic phase to the
low-temperature nematic phase.  For a larger value $\kappa=0.75$, the
isotropic-tetratic transition is still second-order, but now the
tetratic-nematic transition is first-order, with a discontinuous change in
$C_2$.  For $\kappa=1.5$, the two transitions merge into a single first-order
transition directly from isotropic to nematic, with discontinuities in both
$C_2$ and $C_4$, and the tetratic phase does not occur.  Finally, for the
largest value $\kappa=2.25$, the isotropic-nematic transition becomes
second-order; this behavior corresponds to the prediction of 2D Maier-Saupe
theory with a simple $\cos2\theta_{12}$ interaction.

\begin{figure}
\includegraphics{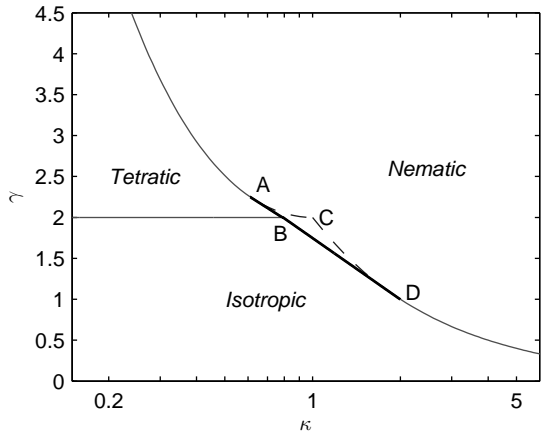}
\caption{\label{fig:phase_diagram_theory} Phase diagram of the model in terms
of $\gamma$ (inverse temperature) and $\kappa$ (two-fold distortion in the
interaction).  The grey solid lines represent second-order transitions, and
the dark solid lines are first-order transitions.  The dashed lines indicate
the extrapolated second-order transitions, which give the cooling limits of
the metastable phases.  Point B (0.79,2) is the triple point, and A (0.61,2.2)
and D (2,1) are the two tricritical points. Point C (1,2) is the intersection
of the extrapolated second-order transitions.}
\end{figure}

The numerical mean-field results are summarized in the phase diagram of
Fig.~\ref{fig:phase_diagram_theory}.  The system has an isotropic phase at low
$\gamma$ (high temperature) and a nematic phase at high $\gamma$ (low
temperature).  It also has an intermediate tetratic phase, as long as the
symmetry-breaking $\kappa$ is sufficiently small.  The temperature range of
the tetratic phase is very large for small $\kappa$, then it decreases as
$\kappa$ increases, and finally vanishes at the triple point B.  In this
mean-field approximation, the isotropic-tetratic transition is always
second-order and independent of $\kappa$.  The tetratic-nematic transition is
second-order for small $\kappa$, then becomes first-order at the tricritical
point A.  The direct isotropic-nematic transition is second-order for large
$\kappa$, then becomes first-order at the tricritical point D.  Point C is the
intersection of the extrapolated second-order transitions, and represents the
limit of metastability of the tetratic phase.

To calculate second-order transitions and special points in the phase diagram,
we minimize the free energy of Eq.~(\ref{eq:free}) analytically.  For this
calculation, we expand the free energy as a power series in the order
parameters $C_2$ and $C_4$, which gives
\begin{eqnarray}
\frac{F}{k_B T}&=&\frac{\gamma\kappa(2-\gamma\kappa)}{8} C_2^2
+\frac{\gamma(2-\gamma)}{4} C_4^2
-\frac{\kappa^2 \gamma ^3}{8} C_2^2 C_4\nonumber \\
& &+\frac{\kappa^4 \gamma ^4}{64} C_2^4
+\frac{\gamma^4}{64} C_4^4 +\ldots.
\label{eq:landau_f}
\end{eqnarray}
Note that this expression is exactly what would be expected in a Landau
expansion based on symmetry; it is always an even function in $C_2$, but it is
an even function of $C_4$ only when $C_2=0$.

To find the isotropic-tetratic transition, we set $C_2=0$ in the expansion,
because this order parameter vanishes in both of those phases.  The
second-order isotropic-tetratic transition then occurs when the coefficient of
$C_4^2$ passes through 0.  Hence, the transition is at
\begin{equation}
\label{eq:isotropictetratic}
\gamma=2,
\end{equation}
independent of $\kappa$.

For the second-order isotropic-nematic transition, we see that the isotropic
phase becomes unstable when $\partial^2 F/\partial C_2^2 =
\partial^2 F/\partial C_4^2 = \partial^2 F/\partial C_2 \partial C_4 = 0$,
all evaluated at $C_2=C_4=0$.  These equations have two solutions, one of
which corresponds to the isotropic-tetratic transition found above.  The other
solution, representing the isotropic-nematic transition, is
\begin{equation}
\label{eq:isotropicnematic}
\gamma=\frac{2}{\kappa}.
\end{equation}
On the nematic side of this transition, we find
$C_4=\kappa^2 \gamma^2 C_2^2/[4(2-\gamma)]$; i.e. the order parameters $C_2$
and $C_4$ increase with different critical exponents.  We substitute that
relation into the expansion~(\ref{eq:landau_f}) to obtain an effective free
energy in terms of $C_2$ alone,
\begin{equation}
\frac{F_{\textrm{eff}}}{k_B T}=\frac{\gamma\kappa(2-\gamma\kappa)}{8} C_2^2
+\frac{\gamma^4 \kappa^4 (1-\gamma)}{32(2-\gamma)}C_2^4 + \ldots.
\end{equation}
The tricritical point D occurs when the coefficients of \emph{both} $C_2^2$
and $C_2^4$ vanish in this expansion, which is at $\gamma=1$ and $\kappa=2$.

\begin{figure}
\includegraphics{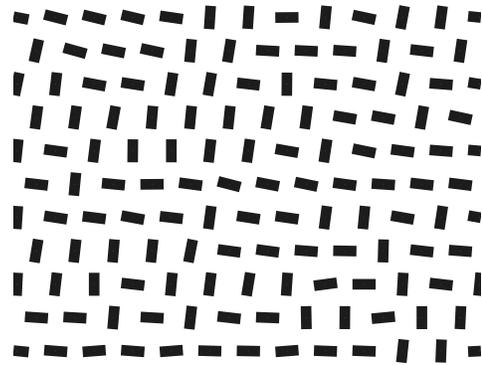}
\caption{\label{fig:lattice} A snapshot of the spins on a triangular lattice
in the tetratic phase.  The shape of the rectangles is just a schematic
illustration of the symmetry of their interaction.}
\end{figure}

For the second-order tetratic-nematic transition, we cannot use the expansion
of Eq.~(\ref{eq:landau_f}) because $C_4$ is not necessarily small; instead we
return to the free energy of Eq.~(\ref{eq:free}).  Anywhere in the tetratic
phase we must have $\partial F/\partial C_4 = 0$, which implies
\begin{equation}
\label{eq:tetratic}
C_4=\frac{I_1(C_4\gamma)}{I_0(C_4\gamma)},
\end{equation}
where $I_0$ and $I_1$ are modified Bessel functions of the first kind.  At the
tetratic-nematic transition, we also have $\partial^2 F/\partial C_2^2=0$,
evaluated at $C_2=0$, which implies
\begin{equation}
2-\gamma\kappa=\frac{\gamma\kappa I_1(C_4\gamma)}{I_0(C_4\gamma)}.
\end{equation}
These two equations implicitly determine the second-order tetratic-nematic
transition line shown in Fig.~\ref{fig:phase_diagram_theory}.  To find the
tricritical point A, we expand the free energy in powers of $C_2$, for $C_4$
satisfying Eq.~(\ref{eq:tetratic}), and we require that the coefficients of
$C_2^2$ and $C_2^4$ both vanish.  As a result, the tricritical point A occurs
at $\gamma=2.2496$, $\kappa=0.6116$ and $C_4=0.4535$.

\begin{figure*}
(a)\subfigure{\includegraphics{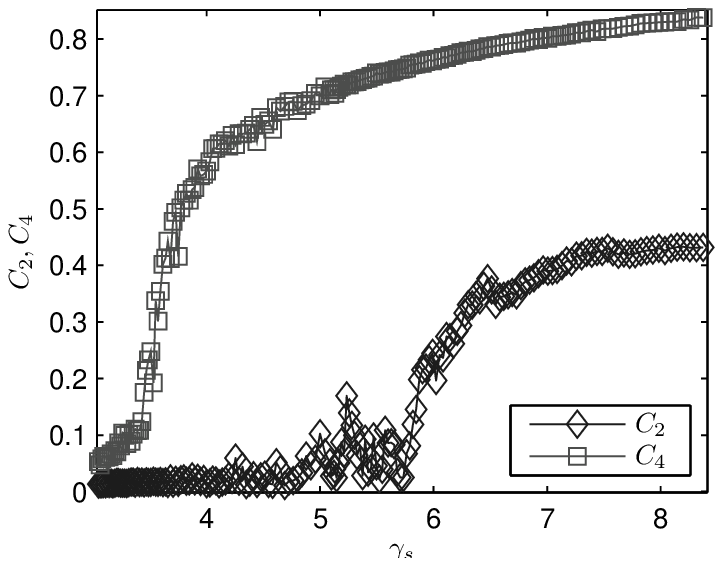}}
(b)\subfigure{\includegraphics{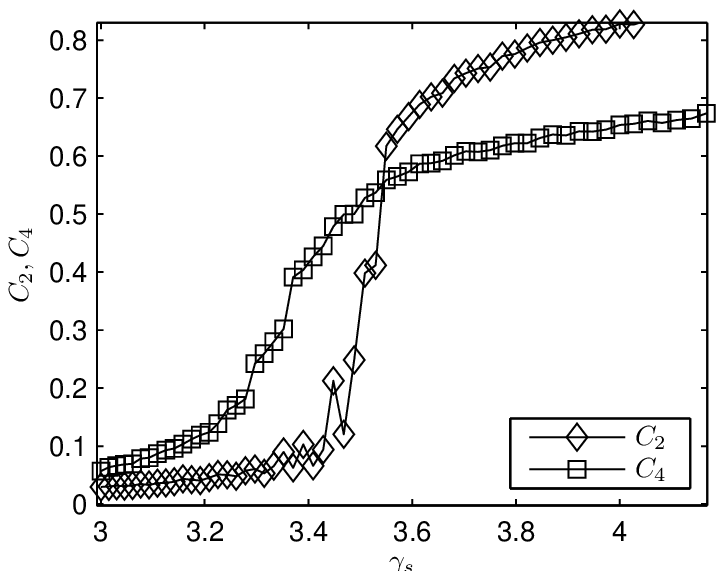}}
(c)\subfigure{\includegraphics{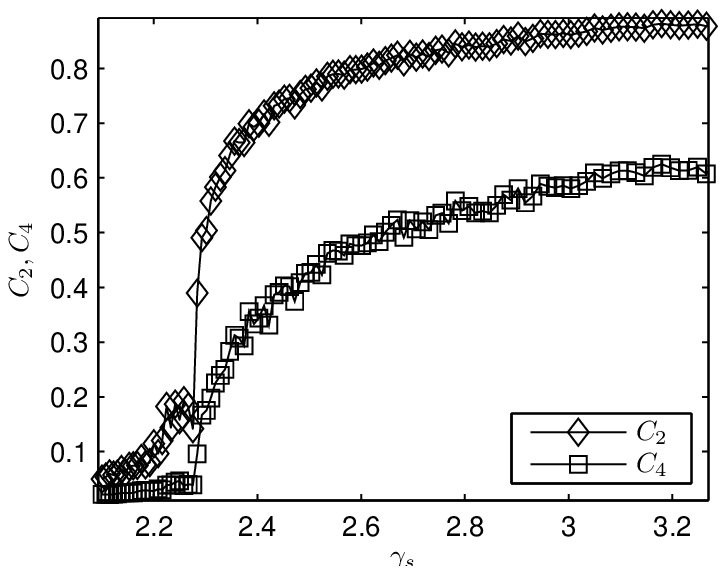}}
(d)\subfigure{\includegraphics{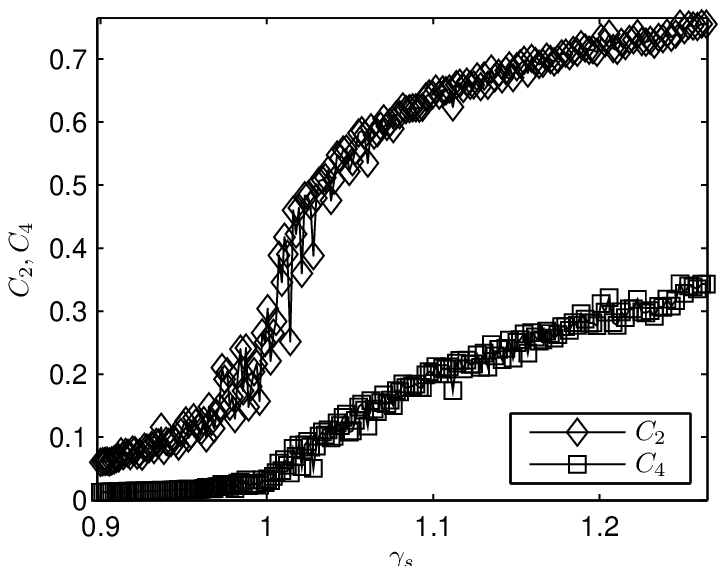}}
\caption{\label{fig:Csgks} Simulation results for the order parameters $C_2$
and $C_4$ as functions of $\gamma$ (inverse temperature), for several values
of $\kappa$ (two-fold distortion in the interaction):  (a)~$\kappa=0.3$.
(b)~$\kappa=0.5$.  (c)~$\kappa=1$.  (d)~$\kappa=3$.}
\end{figure*}

The first-order transition lines in the phase diagram cannot be calculated
analytically; instead they are determined by numerical minimization of the
free energy.  The triple point B occurs where the first-order transition lines
intersect the second-order isotropic-tetratic transition of
Eq.~(\ref{eq:isotropictetratic}).  This point is found numerically at
$\gamma=2$ and $\kappa=0.79$.

Point C is the intersection of the extrapolated second-order transitions of
Eqs.~(\ref{eq:isotropictetratic}) and~(\ref{eq:isotropicnematic}), which
occurs at $\gamma=2$ and $\kappa=1$.  It represents the highest value of the
symmetry breaking $\kappa$ where the tetratic phase can even be metastable,
beyond the the triple point B where it ceases to be a stable phase.

\section{\label{sec:simulation}Monte Carlo simulations}

So far, the calculations presented in this paper have all used mean-field
theory.  Of course, mean-field theory is an approximation, which tends to
exaggerate the tendency toward ordered phases.  In order to assess the
validity of mean-field theory, we perform Monte Carlo simulations for a
lattice model of the same system.  In this lattice model, we use the
Hamiltonian
\begin{equation}
H=-J\sum_{\left<i,j\right>}
\left(\kappa\cos[2(\theta_i-\theta_j)]+\cos[4(\theta_i-\theta_j)]\right),
\label{eq:H_simulation}
\end{equation}
summed over nearest-neighbor sites $i$ and $j$ on a 2D triangular lattice, as
shown in Fig.~\ref{fig:lattice}.  This lattice Hamiltonian corresponds to the
model presented in the previous sections if we take the parameter
$\gamma=6J/(k_B T)$, because each lattice site interacts with six nearest
neighbors.

We simulate this model on a lattice of size $100\times 100$ with periodic
boundary conditions, using the standard Metropolis algorithm~\cite{
metropolis:1087}. On each lattice
site, the spin is described by an orientation angle $\theta$.  In each trial
Monte Carlo step, a spin is chosen randomly, its orientation is changed
slightly, and the resulting change in the energy $\Delta E$ is calculated. If
energy decreases, the change is definitely accepted. If not, the change is
accepted with a probability of $\exp\left[-\Delta E/(k_B T)\right]$. Usually,
for a constant temperature, each Monte Carlo cycle of the simulation consists
of 10000 trial steps, and 50000 cycles are used for each temperature. However,
near phase transitions, especially near first-order transitions, additional
Monte Carlo cycles are used to eliminate metastable states. The phase diagram
is calculated by cooling the system from high temperature with decreasing the
temperature in steps of 0.01, or steps of 0.005 near phase transitions. During
the last half of the simulation cycles, the order parameters are calculated
and time-averaged.

To calculate the nematic order parameter $C_2$, we use the 2D nematic order
tensor
\begin{equation}
Q_{\alpha \beta}=2\left(\left<n_\alpha n_\beta\right>
-\left<n_\alpha n_\beta\right>_\textrm{iso}\right),
\end{equation}
averaged over all lattice sites.  Here, ${\bf n}=(\cos\theta,\sin\theta)$ is
the unit vector representing each spin, and
$\left<n_\alpha n_\beta\right>_\textrm{iso}=\frac{1}{2}\delta_{\alpha \beta}$
is the average in the isotropic phase.  The positive eigenvalue of this tensor
is $C_2$.

For the tetratic order parameters $C_4$, we use the generalized tensor method
of Zheng and Palffy-Muhoray~\cite{PeterPalffy--1}.  We consider the
fourth-order tetratic order tensor
\begin{equation}
T_{\alpha \beta \gamma \delta}=
4\left(\left<n_\alpha n_\beta n_\gamma n_\delta \right>
-\left<n_\alpha n_\beta n_\gamma n_\delta \right>_\textrm{iso}\right),
\end{equation}
averaged over all lattice sites.  Here, we are subtracting off the isotropic
average $\left<n_\alpha n_\beta n_\gamma n_\delta \right>_\textrm{iso}=
\frac{1}{8}(\delta_{\alpha\beta}\delta_{\gamma\delta}
+\delta_{\alpha\gamma}\delta_{\beta\delta}
+\delta_{\alpha\delta}\delta_{\beta\gamma})$.  To calculate the eigenvalues,
we unfold this fourth-order tensor into a second-order tensor ($4 \times 4$
matrix), which we diagonalize using standard methods.  The four eigenvalues
are $0$, $-C_4$, $\frac{1}{2}\left(C_4-(16C_2^2+C_4^2)^{1/2}\right)$, and
$\frac{1}{2}\left(C_4+(16C_2^2+C_4^2)^{1/2}\right)$.  (In the tetratic phase,
with $C_2=0$, they reduce to $0$, $-C_4$, $+C_4$, and $0$.)  Thus, using the
previously calculated value of $C_2$, we can extract $C_4$.

Figure~\ref{fig:Csgks} shows the simulation results for the order parameters
$C_2$ and $C_4$ as functions of $\gamma$, for several values of $\kappa$.
These results are quite simular to the numerical mean-field results of
Fig.~\ref{fig:Csgkt}, although the quantitative values of $\gamma$, $\kappa$,
and the order parameters are somewhat different.  For a small symmetry
breaking $\kappa=0.3$, there are two second-order phase transitions.  The
order parameter $C_4$ increases continuously at the high-temperature
isotropic-tetratic transition, and $C_2$ increases continuously at the
lower-temperature tetratic-nematic transition.  For a slightly larger value of
$\kappa=0.5$, the tetratic-nematic transition becomes first-order, with an
apparently discontinuous increase in $C_2$ (within the precision of the
simulation).  For $\kappa=1$, the intermediate tetratic phase disappears, and
there is just a single first-order isotropic-nematic transition, with
apparently discontinuous increases in both $C_2$ and $C_4$.  Finally, for the
largest value $\kappa=3$, the isotropic-nematic transition becomes
second-order, with continuous increases in both $C_2$ and $C_4$.

\begin{figure}
\includegraphics{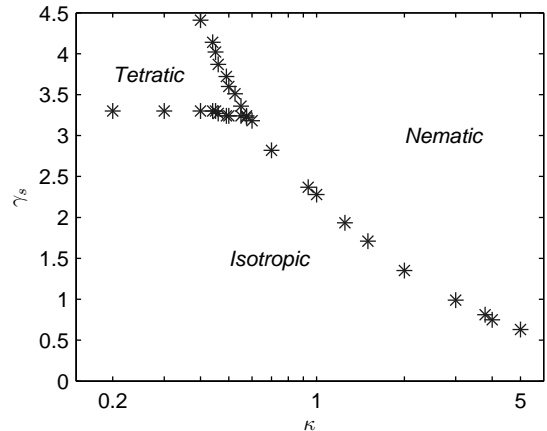}
\caption{\label{fig:phase_diagram_simulation} Simulation results for the phase
diagram in terms of $\gamma$ (inverse temperature) and $\kappa$ (two-fold
distortion in the interaction).  The triple point is at approximately
$\gamma=3.2$ and $\kappa=0.60$.}
\end{figure}

The simulation results are summarized in the phase diagram of
Fig.~\ref{fig:phase_diagram_simulation}.  This phase diagram shows a
high-temperature isotropic phase, an intermediate tetratic phase, and a
low-temperature nematic phase.  The temperature range of the tetratic phase is
very large when the symmetry breaking $\kappa$ is small, then it decreases as
$\kappa$ increases, and eventually vanishes at the triple point, which is
approximately given by $\gamma=3.2$ and $\kappa=0.60$.
Compared with the mean-field phase diagram of
Fig.~\ref{fig:phase_diagram_theory}, the simulation results show the
transitions at lower temperature (higher $\gamma$) than in mean-field theory.
This difference is reasonable because mean-field theory always exaggerates the
tendency toward ordered phases.

\section{\label{sec:conclusions}Conclusions}

In this paper, we propose a model for the statistical mechanics of particles
with almost-four-fold symmetry.  In contrast with earlier work on particles
with a hard-core excluded-volume interaction, we consider particles with a
soft interaction, analogous to Maier-Saupe theory of nematic liquid crystals.
We investigate this model through two complementary techniques, mean-field
calculations and Monte Carlo simulations.  Both of these techniques predict a
phase diagram with a low-temperature nematic phase, an intermediate tetratic
phase, and a high-temperature isotropic phase.  They make consistent
predictions for the order of the transitions and the temperature dependence of
the order parameters, although they do not agree in all quantitative details.

The main result of this study is that the tetratic phase can exist up to a
surprisingly high value of the symmetry breaking $\kappa$ in the microscopic
interaction.  We find the maximum $\kappa=0.79$ in mean-field theory, or 0.60
in Monte Carlo simulations.  Even taking the lower Monte Carlo value, this
implies that the interaction in the parallel direction $(1+\kappa)$ can be
about four times larger than the interaction in the perpendicular direction
$(1-\kappa)$.  Hence, the tetratic phase can form even for particles with
quite a substantial two-fold component in the interaction energy, i.e. for
fairly rod-like particles.

It is interesting to note that our phase diagram is quite similar to the phase
diagram found by density-functional theory for hard rectangles; see Fig.~3 of
Ref.~\cite{Martinez-Rat05}.  In that theory, the phase diagram shows
isotropic, tetratic, and nematic phases, and the tetratic phase can exist for
rectangles with aspect ratio of up to 2.21:1.  That theory shows the same
arrangement of the phases, and even the same first- and second-order phase
transitions, with tricritical points on the isotropic-nematic and
tetratic-nematic transition lines.  This phase diagram seems to be a generic
feature of particles with four-fold symmetry broken down to two-fold. Thus, we
can expect to see tetratic phases in 2D experiments and simulations, even if
the particles are moderately extended.

\begin{acknowledgments}
We would like to thank R. L. B. Selinger and F. Ye for many helpful
discussions.  This work was supported by the National Science Foundation
through Grant DMR-0605889.
\end{acknowledgments}

\newpage
\bibliography{PRE_tetratic}

\end{document}